\documentclass{PoS}
\usepackage{mathtools}
\usepackage{amsmath,amstext}
\usepackage{wrapfig}
\usepackage{subcaption}
\usepackage[rawfloats=true]{floatrow}
\usepackage{url}
\title{Simulation and Reconstruction Study of a Future Surface
Scintillator Array at the IceCube Neutrino Observatory}

\ShortTitle{Simulation and Reconstruction of Scintillator Array at IceCube}

\author{The IceCube Collaboration\footnote{For collaboration list, see PoS(ICRC2019) 1177.}\\
{\itshape \href{http://icecube.wisc.edu/collaboration/authors/icrc19_icecube}{http://icecube.wisc.edu/collaboration/authors/icrc19\_icecube}}\\
E-mail: \email{agnieszka.leszczynska@kit.edu, matthias.plum@marquette.edu}
        }

\abstract{
The IceCube Neutrino Observatory at the South Pole is a multi-component particle detector consisting of the IceTop surface array and the deep in-ice IceCube array. The foreseen enhancement of the surface instrumentation will consist of plastic scintillator panels read out by silicon photomultipliers. This additional detector component will calibrate the effect of snow accumulation on the IceTop tanks, improve the measurement of cosmic rays, and enhance the atmospheric background rejection for the high-energy astrophysical neutrino detection.

Two scintillator prototype stations were deployed at IceTop in the austral summer of 2017/18 to test the detector design and have started taking data. In order to understand the properties of the scintillator panel response a detailed Geant4 simulation of a single detector, including the photon propagation and simulated SiPM response, is being developed and parameterized. We investigate the capabilities of the IceTop upgrade with an optimized layout of the new detectors and the accuracy of the reconstruction. We will present the details of the simulation and reconstruction studies for the proposed IceTop enhancement and report the capabilities of the combined installation.

\vspace{4mm}
{\bfseries Corresponding authors:}
\speaker{ Agnieszka Leszczy{\'n}ska}$^{1}$, Matthias Plum$^{2}$\\
{$^{1}$ \itshape Karlsruhe Institute of Technology, Institut f{\"u}r Kernphysik, D-76021 Karlsruhe, Germany}\\
{$^{2}$ \itshape Department of Physics, Marquette University, Milwaukee, WI, 53201, USA}
}

\FullConference{36th International Cosmic Ray Conference -ICRC2019-\\
		July 24th - August 1st, 2019\\
		Madison, WI, U.S.A.}

\begin{document}
\section{IceTop Surface Enhancement}\label{sec:intro}
\begin{figure}[ht]
\centering
\includegraphics[width=33pc]{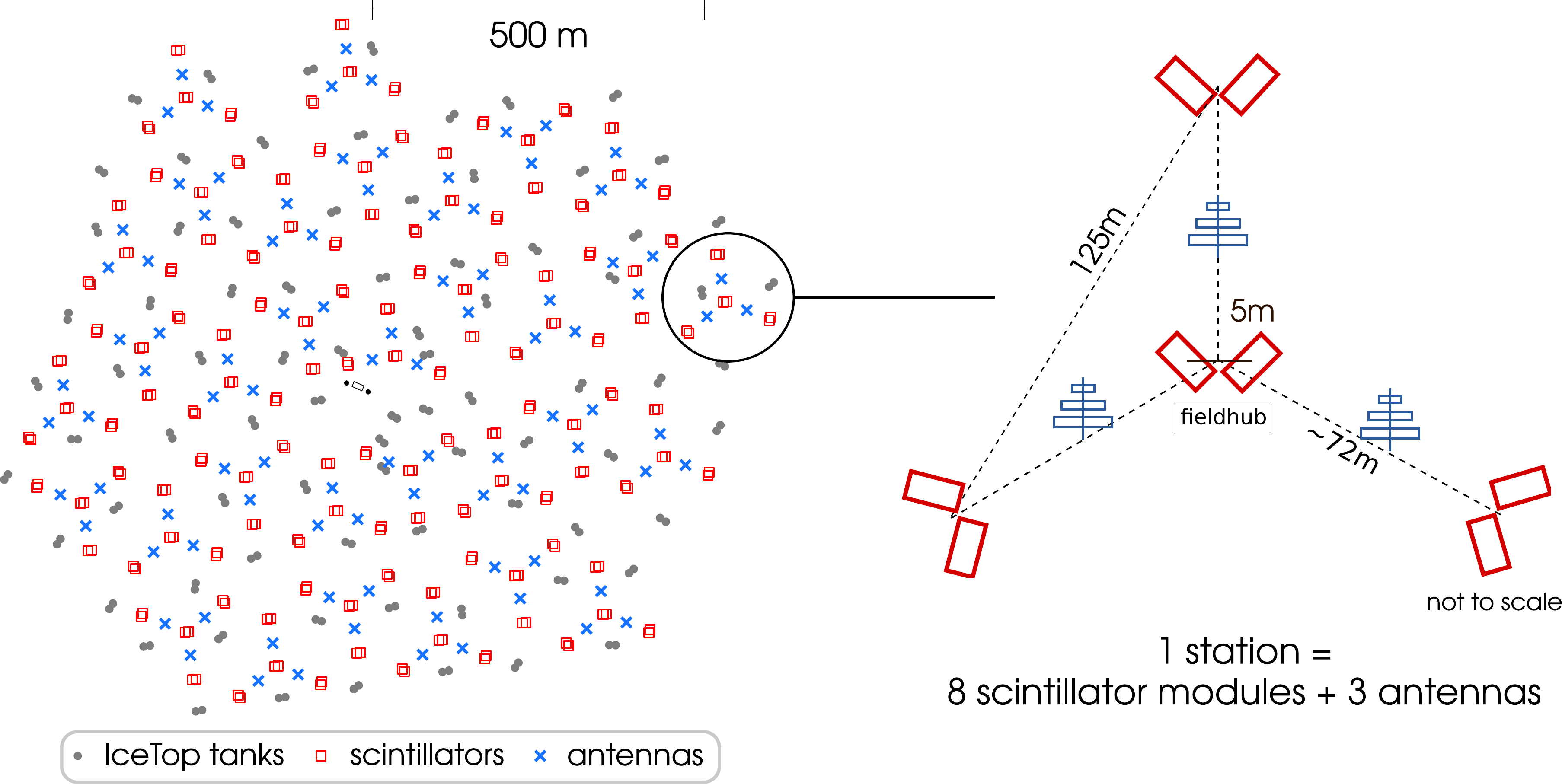}
\caption{\label{icetopext}The scheme of the IceTop enhancement. \textit{Left:} Optimized layout of the enhanced array. Red squares indicate planned positions of the scintillation detectors, blue crosses of antennas and gray dots show positions of the current IceTop tanks.  \textit{Right:} Structure of one hybrid station.}
\end{figure}

The surface array at the IceCube Neutrino Observatory, called IceTop, is a cosmic ray detector measuring extensive air showers initiated by primary particles with energies from PeV to EeV, covering the transition region from galactic to extra-galactic sources~\cite{Aartsen:2013IT}. Due to constant snow accumulation, around 20\,cm$/$year on average~\cite{Rawlins:2015ICRC}, and too sparse sampling of the shower footprint, the energy threshold remains too high for a combined investigation of the cosmic-ray spectrum and composition at the energy range of the knee feature. The planned enhancement of IceTop~\cite{haungs} foresees, among other upgrades, a deployment of 256 plastic scintillation detectors which will enable extending the energy range by doubling the current sensitive area and increasing the number of detectors. The layout of the panels is presented in Figure~\ref{icetopext} with 32 stations of 8 detectors each. This particular design was optimized for trenching-length reduction and uniform distribution of the scintillator panels among the IceTop tanks. An advantage of installing a detector of a new type is the ability to make complementary measurements of air-shower components to the Cherenkov tanks, which increases the potential for the cosmic-ray species determination. Two prototype scintillator stations were deployed in 2018 at the South Pole and have demonstrated an ability to detect air showers~\cite{Kauer:2019icrcw}.

Additionally, every scintillator station will be equipped with 3 radio antennas. Such an extension of the standard IceTop array will make it a novel type of hybrid cosmic-ray detector. The radio detection technique is well established and has proven its capabilities of, among other things, mass separation~\cite{Holt2019} and  measurement of inclined air showers, which opens new science cases~\cite{Schoeder:2019icrcw}.

In this work we present the simulation and reconstruction procedures of the scintillator part of this upgrade as well as a preliminary estimation of its  capabilities in combination with the IceTop tanks. 

\section{Scintillation detector simulation}\label{sec:sim}

An understanding of the response of the detector array can be achieved only with comprehensive simulation studies. The detector response of a complete, detailed design was simulated using the Geant4.10 toolkit ~\cite{Agostinelli:2002hh}. Each detector panel consists of 16 Fermilab scintillator bars~\cite{fermilab_scint} coated with a reflective TiO$_2$ dye, comprising 1.5\,m$^2$ (1.875\,m\,$\times$\,0.8\,m) active area. A 1\,mm diameter wavelength-shifting fiber, Kuraray\,\footnote{\url{http://kuraraypsf.jp/pdf/all.pdf}} Y11(300), is connected to a silicon photomultiplier~(SiPM), Hamamatsu\,\footnote{\url{https://www.hamamatsu.com/us/en/product/type/S13360-6025PE/index.html}} 13360-6025, with optical gel. The sensitive part of the panel is supported by styrofoam and plywood and is put in an aluminum casing ~\cite{Collaboration:2017tdy}.

\begin{figure}[ht]
  \floatsetup{heightadjust=all, valign=c}
  \begin{floatrow}
    \ffigbox{%
      \includegraphics[width=0.55\textwidth]{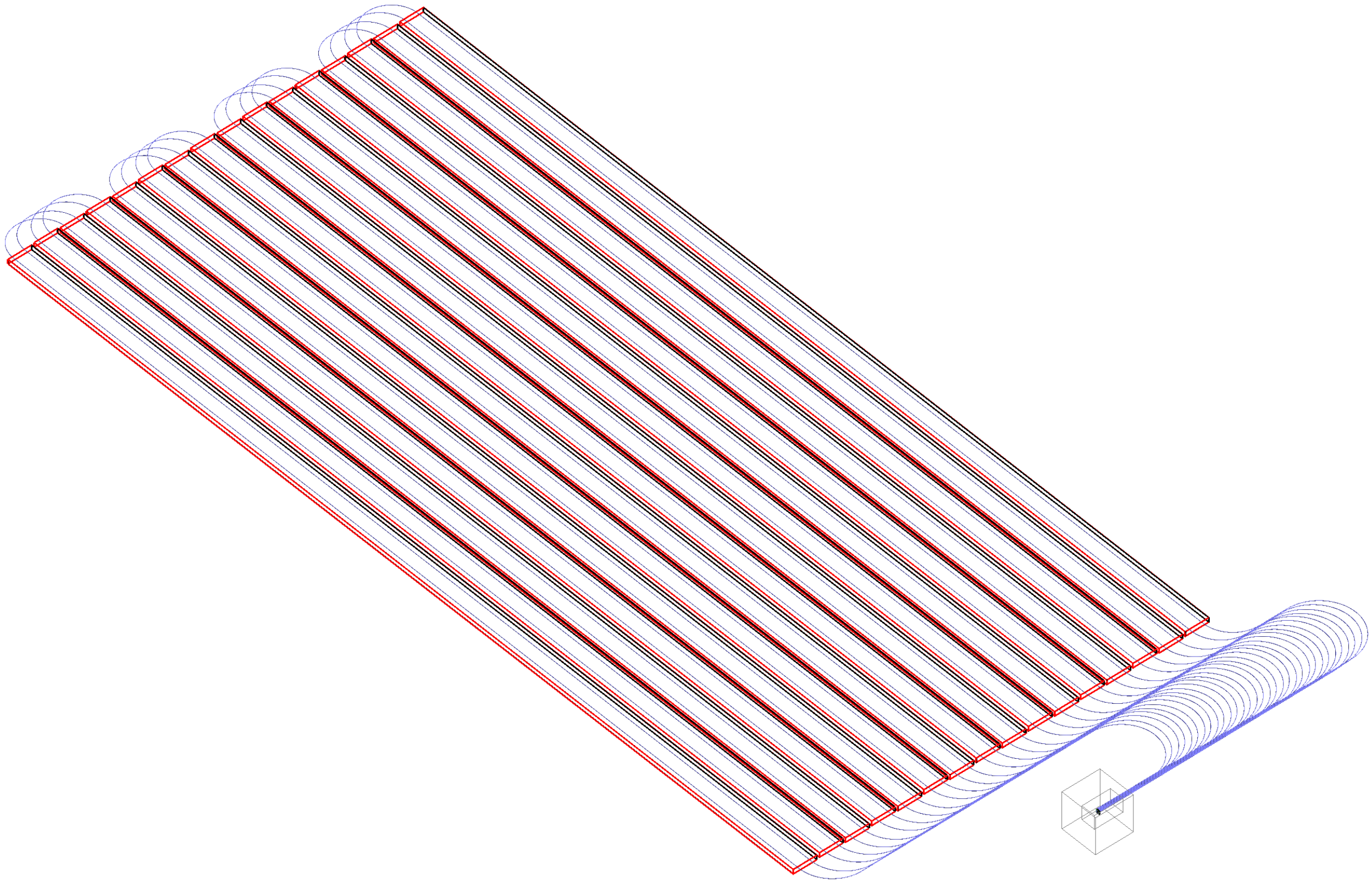}
    }{%
      \captionof{figure}{Scintillation detector model including multi-cladding wavelength shifting fibers (bars: 1875\,mm\,$\times$\,50\,mm\,$\times$\,10\,mm; fiber: 1\,mm\,\O).}%
      \label{geant4_station_model}%
    }
    \ffigbox{%
      \includegraphics[width=0.34\textwidth]{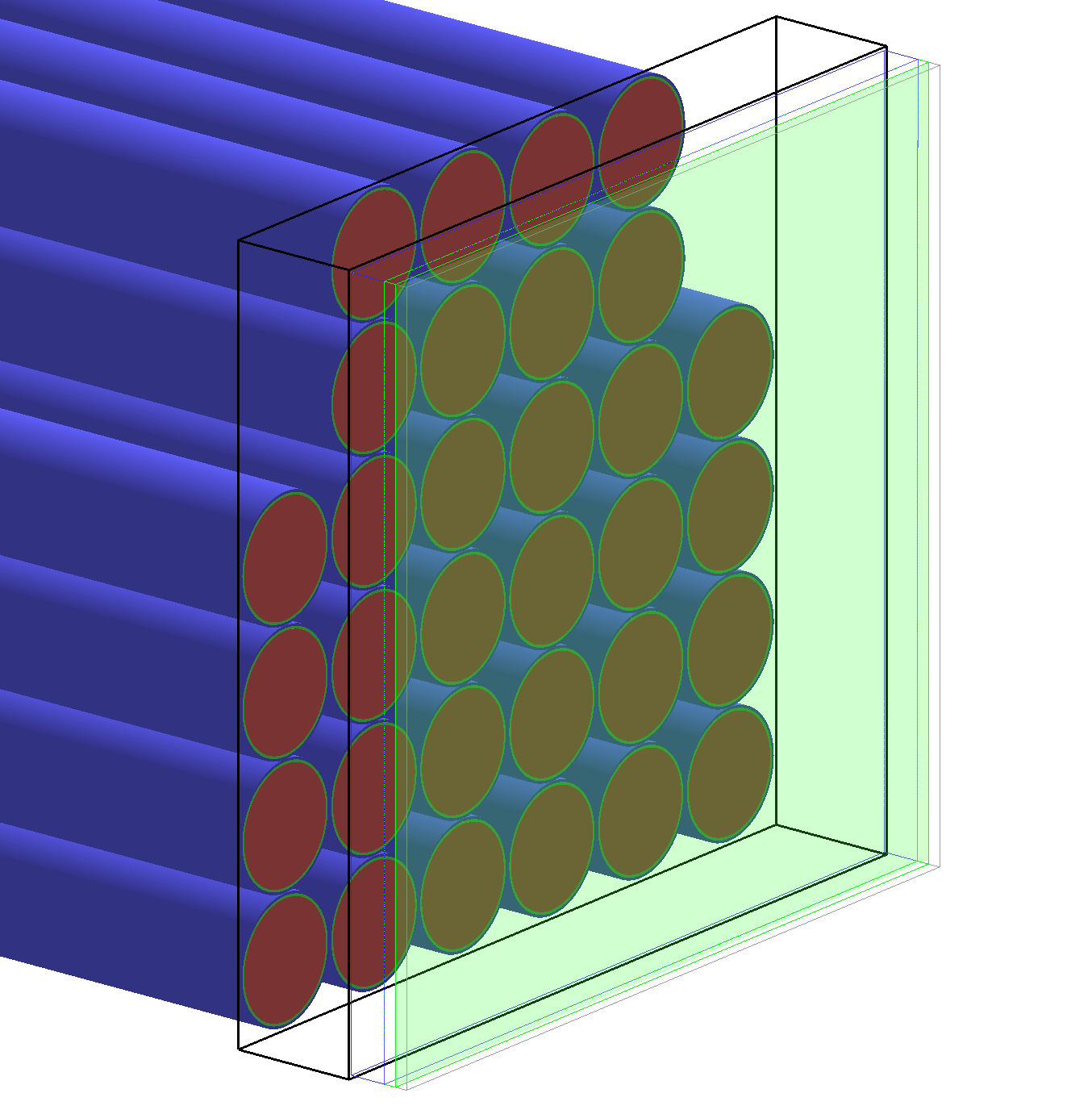}
    }{%
      \captionof{figure}{Model of SiPM and multi cladding fiber coupling. SiPM sensitive area is shown in green, SiPM coating in blue and the  optical gel in black.}%
      \label{geant4_sipm_model}%
    }
  \end{floatrow}
\end{figure}

In Figures \ref{geant4_station_model} and \ref{geant4_sipm_model} the final simulated model view is shown, including the scintillator bars with reflective coating, the multi-cladding fibers~\cite{Dietz_Laursonn_2017}, and the SiPM~\cite{NIGGEMANN2015344} which includes an electronic noise simulation with thermal noise, afterpulses, and crosstalk. 

\begin{figure}[ht]
\centering
\includegraphics[width=27pc]{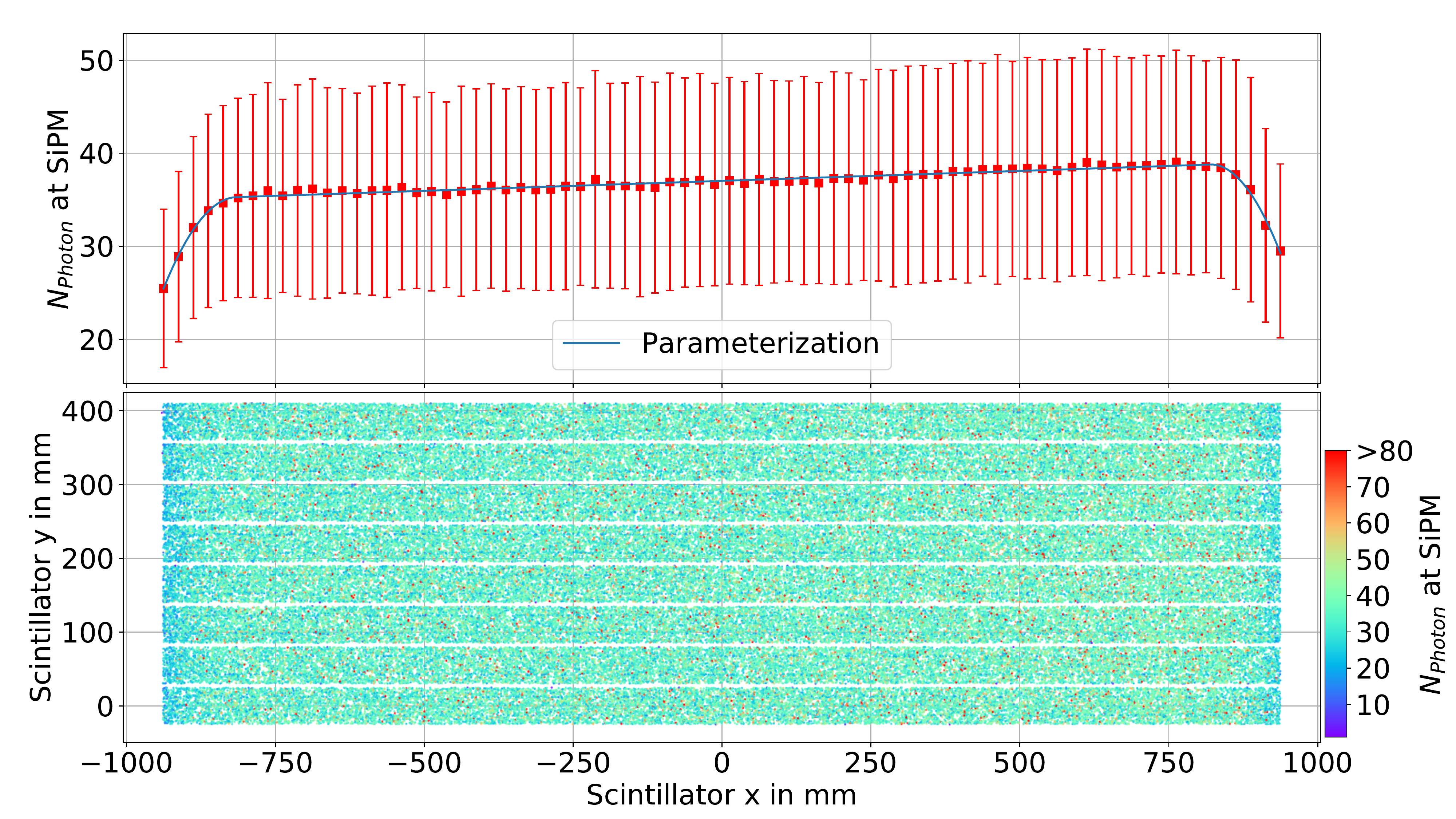}
\caption{\label{geant4_muon_hits} \textit{Upper:} Mean number of photons in each slice of the lower plot. The uncertainty represents the standard deviation. \textit{Lower:} Number of photons vs. muon hit position along the detector. }
\centering
\includegraphics[width=27pc]{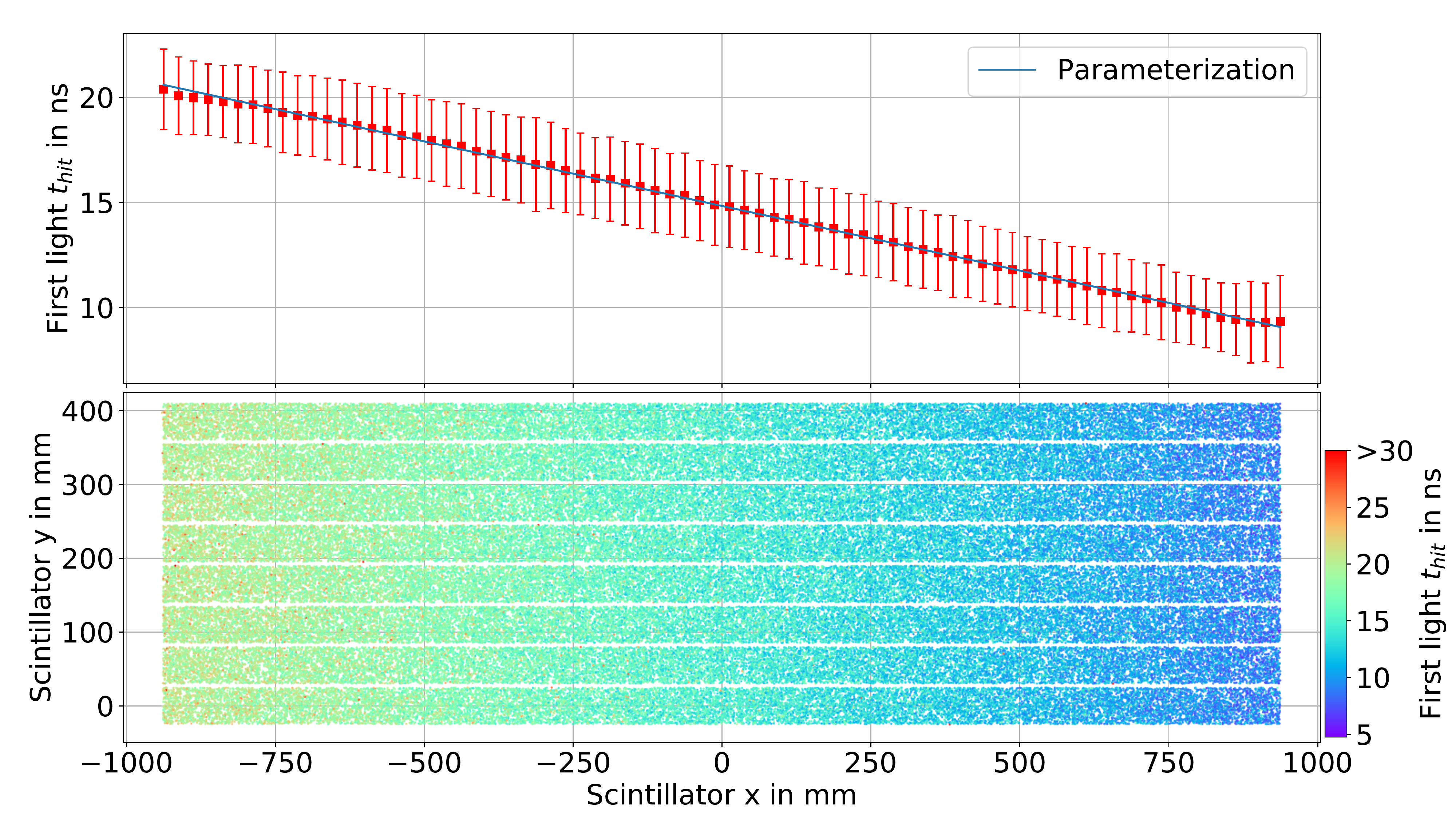}
\caption{\label{geant4_muon_hits_time}  \textit{Upper}: Mean time of first photon in each slice of the lower plot. The uncertainty represents the standard deviation. \textit{Lower:} Scatter plot of the first detected light hit time vs. muon hit position along the detector.
}
\end{figure}

A description of the single detector response was obtained using simulations of 3\,GeV vertical muons. The strength of the simulated SiPM signal and the time delay are shown in Figure~\ref{geant4_muon_hits} and~\ref{geant4_muon_hits_time} respectively. The results were crosschecked with muon-tower measurements~\cite{Collaboration:2017tdy} and they agree very well within the construction-dependent fluctuations. As expected, the efficiency of photon detection by the panel increases towards the detector readout (on the right side in Figure~\ref{geant4_muon_hits} and~\ref{geant4_muon_hits_time}) and is reduced at both of the bar ends due to photons escaping from the scintillator bar. The efficiency, the timing of the light detection, and the shape of the simulated waveforms were parameterized and included in the larger scale simulation. This significantly reduces the computational requirements for creating a large library of simulated air showers.

In large-scale simulations, the number of scintillation photons generated in a panel is modified by the parameterized efficiency, which depends on the interaction position of the incoming secondary particle. This value is converted to the unit of vertical equivalent muon (VEM)~\cite{Leszczy_ska_2019}  and convoluted with the SiPM noise pulses. In order to simulate the real electronic system, a simple trigger is applied. If at least 4 photo-electrons~(PE) are detected, all PEs within a 200\,ns window are summed-up and an additional cut of $\geq$\,0.5\,VEM is applied, accounting for possible electronic losses. 

\section{Scintillator array reconstruction}\label{sec:reco}

To evaluate the performance of the scintillator array, a library of simulated extensive air showers was created and analyzed. The air showers were simulated with CORSIKA v7.6400~\cite{corsika} using FLUKA~\cite{Ferrari_fluka:a,  Bohlen:2014buj} and Sibyll2.3c~\cite{sibyll:2015ICRC} as hadronic interaction models. Proton and iron primaries with energies 10$^{13}$ to 10$^{17}$\,eV were randomly generated from a power law distribution within each energy decade, with zenith angles randomly chosen up to 50$^{\circ}$ from a $\sin\theta\cos\theta$ distribution. Every air shower was sampled 10 times within the central area of the array to account for footprint differences which depend on the impact position on the ground. Secondary particles at 2842\,m\,a.s.l. were read out and propagated through the scintillation detectors. 

The algorithm of air-shower reconstruction includes a negative log-likelihood minimization of the observed signals and times based on the parameterized models. The delay of the signal arrival-time with respect to the shower-front plane can be described by the following function: $\mathrm{\Delta t(r)} = a \exp(-\frac{\mathrm{r}^2}{b^2}) - c \mathrm{r}^2 - d$~\cite{Aartsen:2013IT} with $c$ being a free parameter in the minimization and $a$, $b$, $d$  held constant. 
After evaluating different forms of lateral distribution functions (LDFs)~\cite{Leszczy_ska_2019}, the most robust ones were found to be the LDF currently used by IceTop~\cite{Aartsen:2013IT} and a modified Nishimura-Kamata-Greissen function~\cite{NKGLDF, Leszczy_ska_2019}. Both functions give similar residual values over the typical lateral distance range. In this work we consider only the IceTop-like LDF:
\begin{equation}
\mathrm{S_{\textrm{IceTop}}\left(r\right)=S_{ref} \left( \frac{r}{R_{ref}}\right)^{-\beta-\kappa \log_{10}\left(\frac{r}{R_{ref}}\right)}}
\label{icetopldf}
\end{equation}
$\mathrm{R}_{\mathrm{ref}}$ is a reference distance, $S_{\mathrm{ref}}$ is the signal at this distance and $\kappa$ is a fixed curvature parameter which was tuned using proton and iron showers. $\beta$ is a slope parameter which is free in the minimization. Because the lateral distribution function  can reveal the species (correlated with the LDF slope, $\beta$) and energy (correlated with a signal at a reference distance, S$_{\mathrm{ref}}$) of an incoming cosmic ray, it is of great importance to properly chose a suitable functional form and its parameters. 

\begin{figure}
\centering
\begin{subfigure}{.5\textwidth}
  \centering
  \includegraphics[width=0.95\linewidth]{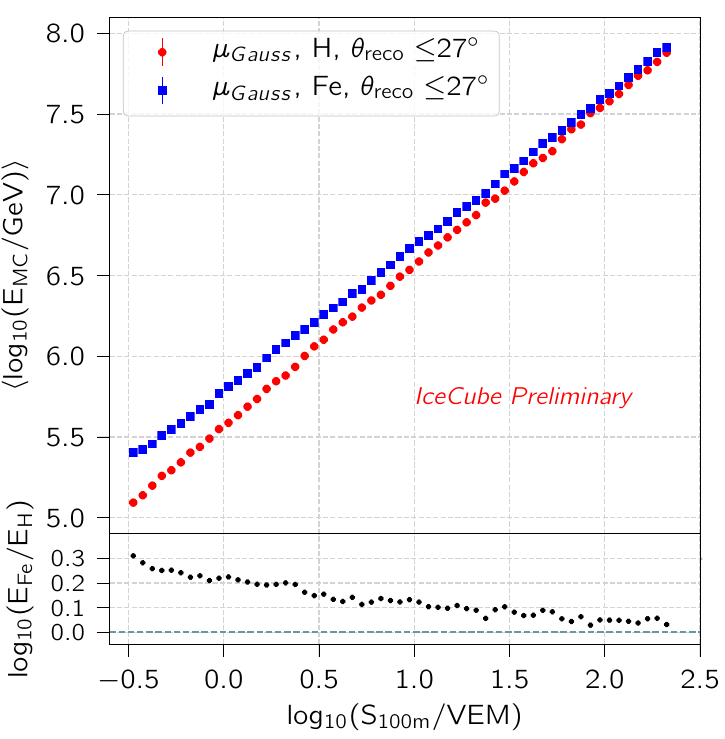}
  \caption{100~m}
  \label{fig:sub1}
\end{subfigure}%
\begin{subfigure}{.5\textwidth}
  \centering
  \includegraphics[width=0.95\linewidth]{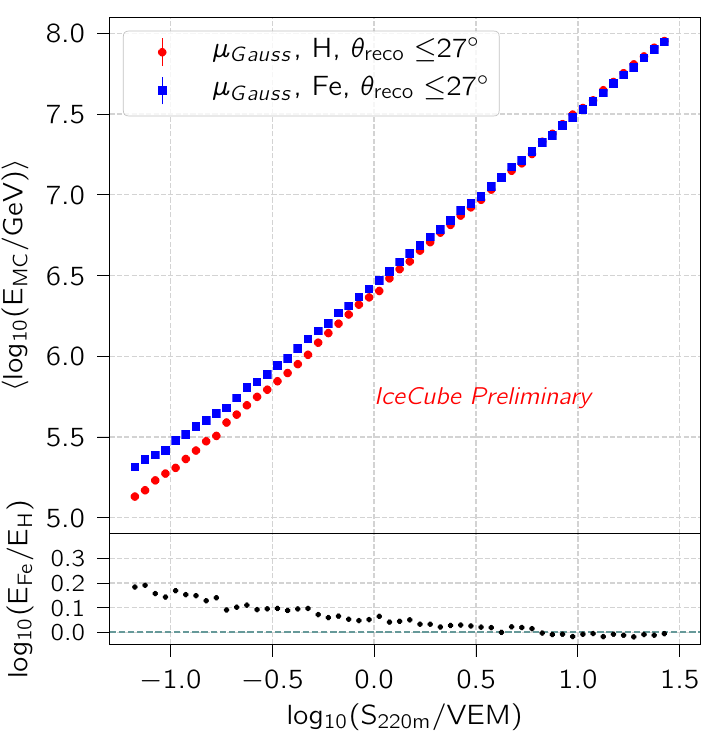}
  \caption{220~m}
  \label{fig:sub2}
\end{subfigure}
\caption{MC primary energy as a function of signal at two reference distances. The markers are the mean of a Gaussian fit of the log$_{10}$(E$_{\mathrm{MC}}/$GeV) values contained in $\log_{10}$(S$_{\mathrm{ref}}/$VEM) bins. Error of the mean values are smaller than the marker size.}
\label{energy100}
\end{figure}
The reconstruction is considered successful when the minimization converges, returning LDF parameters for which $\mathrm{S(r)}$ is monotonically decreasing within the fitted radial range. Only air-showers reconstructed within 400~m from the center of the array are analyzed. 

\begin{wrapfigure}{r}{0.5\textwidth}
\vspace{-15pt}
  \begin{center}
    \includegraphics[width=0.5\textwidth]{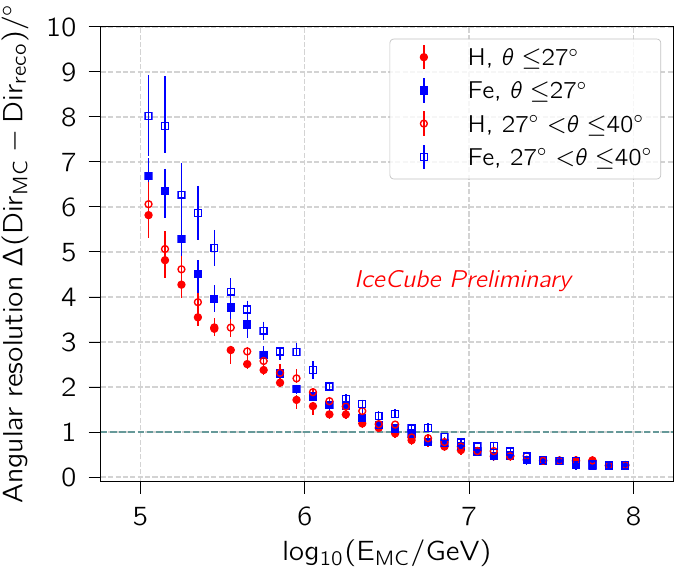}
  \end{center}
  \vspace{-11pt}
  \caption{Angle of incidence reconstruction for the scintillator array. Resolution for a given energy bin is 68th percentile of the binned angular distribution (0.1$^{\circ}$ bin width). The errors were estimated using the bootstrap method and show 95$\%$ confidence intervals.}
    \label{direction}
    \vspace{22pt}
\end{wrapfigure}
$\mathrm{R}_{\mathrm{ref}}$ was chosen to be 220\,m based on the minimization of the covariance between $\beta$ and $\mathrm{S}_{\mathrm{ref}}$ parameters. In Figure~\ref{energy100} the average simulated primary energy as a function of $S_{\mathrm{ref}}$ is shown. The distance of 100\,m was chosen as an example to show that at distances close to the shower axis, the measured signal is highly dependent on the primary mass over the considered energy range. Estimating energy at further distances minimizes this effect, which can be more clearly seen in the lower panels showing the differences between the two primaries. This difference is larger for lower energies and higher zenith angles.
\begin{wrapfigure}{r}{0.51\textwidth}
\vspace{-35.5pt}
  \begin{center}
    \includegraphics[width=0.51\textwidth]{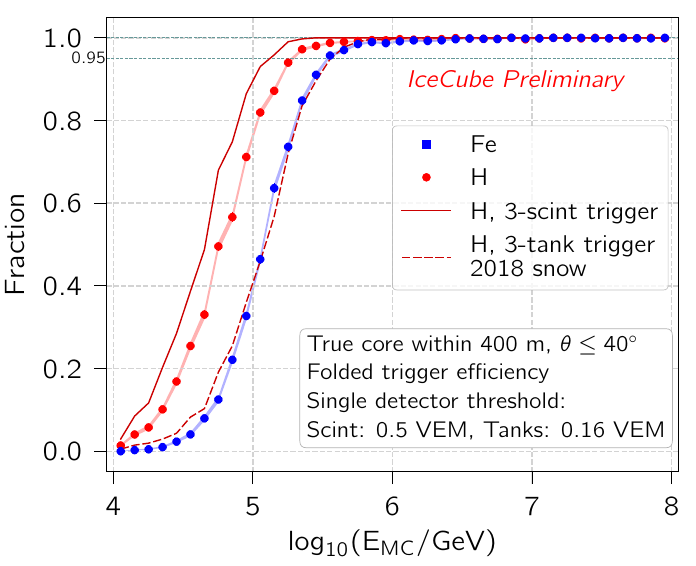}
  \end{center}
  \vspace{-11pt}
  \caption{Reconstruction efficiency for the scintillator array (circles). The red and blue bands visible between the circles are Wilson confidence intervals~\cite{wilson}.  Proton trigger efficiency lines are shown in red for reference (solid for scintillators only, and dashed for tanks only).  Their statistical error bars are small. }
    \label{efficiency}
    \vspace{-10.5pt}
\end{wrapfigure}

The accuracy of the angle of incidence reconstruction versus primary energy is shown in Figure~\ref{direction}. The cosmic ray direction can be reconstructed with a precision of a few degrees below the PeV range. At higher energies, this improves to better than 1$^{\circ}$ and the performance of the scintillator array becomes comparable to that of IceTop~\cite{Rawlins_2016}.

\section{Potential of the enhanced array}\label{sec:outlook}

The trigger condition of the scintillator array requires 3 detectors with a signal of at least 0.5\,VEM within a 1.5\,$\mu$s time window. This will enable lowering the detection threshold, yet will suppress the coincidental noise hits enough to obtain primarily triggers resulting from $\gtrsim$\,100\,TeV air showers. The scintillator array will lower the trigger threshold to $\approx$\,150\,TeV for proton-induced air-showers with zenith angles up to 40$^{\circ}$ and for 95$\%$ efficiency (Figure~\ref{efficiency}).  An even lower trigger threshold can be achieved by combining the two detector arrays. Even though the simulated single-detector trigger threshold is much higher for a scintillation detector than for an IceTop tank, the overall trigger efficiency is much better due to the larger number of detectors and the lack of snow. 
The threshold of the reconstruction efficiency for the simulated events becomes $\approx$\,200\,TeV for proton and $\approx$\,350\,TeV for iron primaries. The scintillator array reconstruction efficiency is shown with circles in Figure~\ref{efficiency}. In this figure, the standard selection of the air showers is applied to the Monte Carlo (MC) values (within 400\,m from the array center and 40$^{\circ}$ of the zenith) to show the difference between the trigger efficiency (dashed and solid lines for tanks and scintillators respectively) and the reconstruction efficiency.

\begin{figure}
\centering
\begin{subfigure}{.5\textwidth}
  \centering
  \includegraphics[width=0.86\linewidth]{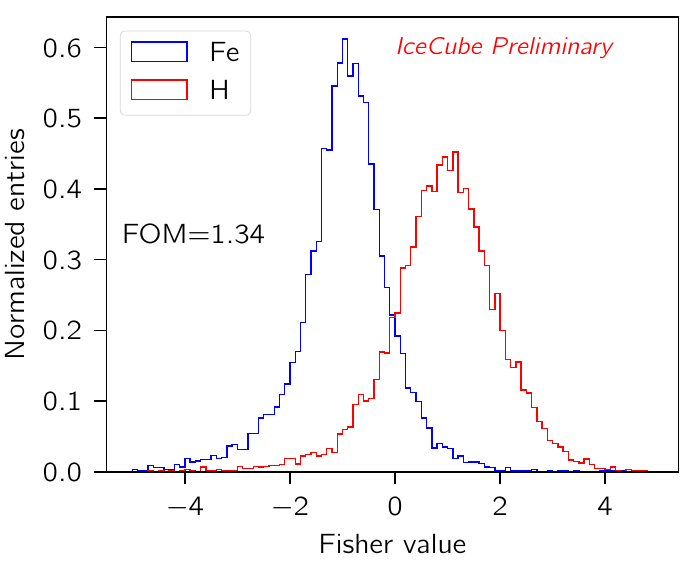}
  \caption{Zenith angular range: 0--27$^{\circ}$ }
  \label{fig:mass1}
\end{subfigure}%
\begin{subfigure}{.5\textwidth}
  \centering
  \includegraphics[width=0.86\linewidth]{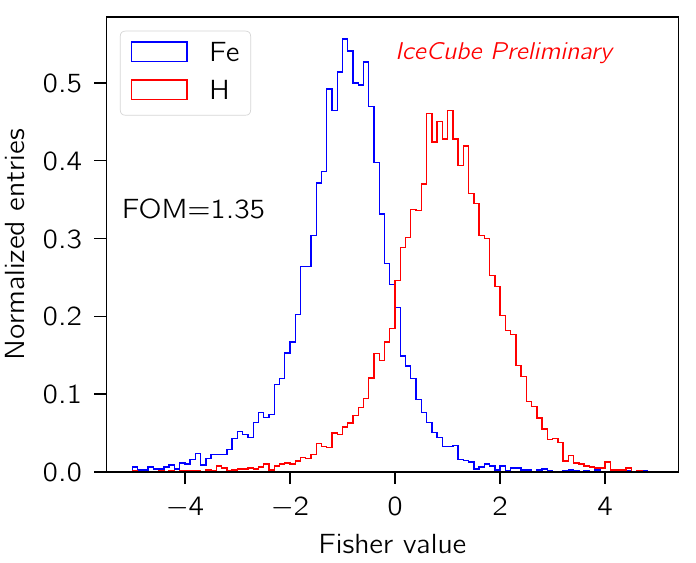}
  \caption{Zenith angular range: 27--40$^{\circ}$ }
  \label{fig:mass2}
\end{subfigure}
\caption{Distribution of Fisher values from a two-parameter analysis for two zenith ranges. The input values for the calculation of the figure of merit (FOM) are taken directly from the histograms.}
\label{pFe_dis}
\end{figure}

Due to the complementary response to secondary particles, scintillation detectors and IceTop tanks together can improve the mass discrimination of cosmic rays. A preliminary analysis of the combined array shows the potential of this approach. With the separate reconstructions for both arrays, two preliminary parameters were selected to differentiate between proton and iron primaries --- the slope of the scintillator LDF, $\beta$, and the ratio of the signal at 200\,m for tank~\cite{Aartsen:2013IT} and scintillator LDFs. The latter parameter takes into account the different response of the two detectors to electrons and muons. Figure \ref{pFe_dis} shows an example of the distributions of Fisher values obtained from these parameters for reconstructed zenith bins of 0--27$^{\circ}$  and 27--40$^{\circ}$  with MC energies from $10^{6}$--$10^{8}$ GeV. 

The separation was obtained with a linear discriminant analysis tool from the scikit-learn package~\cite{scikit-learn}. A figure of merit, FOM = $\lvert \mu_{H}-\mu_{Fe} \rvert  / \sqrt{\sigma_{H}^2 + \sigma_{Fe}^2}$, can be used to express the separation power~\cite{Aab:2016vlz, Holt2019}. For our initial set of parameters this value is above 1.3, which is promising, even without correcting for the primary energy. In order to achieve a better separation power, more detailed studies of the array response will be performed, exploring the possible parameter space for both arrays as well as a combined reconstruction algorithm.

\section{Summary}\label{sec:sum}
The planned enhancement of IceTop will include the installation of scintillator panels and radio antennas. The proposed layout allows a realistic deployment plan without losing sensitivity for the scientific goals. The simulation framework for the new detectors covers the energy deposition and parameterization of the time delay as well as efficiency losses based on detailed Geant4 studies. The analysis of the array response demonstrates a good capability to reconstruct air showers using the scintillation detectors alone. Moreover, combining the scintillator array with IceTop can improve the primary mass discrimination around the knee region. In addition, the future installation of the antenna array will boost this separation even more due to precise measurements of the electromagnetic component. 

The simulations were incorporated in the standard IceCube software. Further studies and validation of the reconstruction and simulation procedure are ongoing as well as a combined reconstruction with multiple detector types within the new reconstruction framework~\cite{Gonzalez:2015rdo, Dvorak:2019icrcw} of the IceCube software.

\bibliographystyle{ICRC}
\bibliography{references}

\end{document}